\documentclass[aps,pra,reprint,amsmath,amssymb,superscriptaddress,onecolumn,longbibliography, notitlepage, nofootinbib]{revtex4-2}

\usepackage{mathtools}
\usepackage{siunitx}
\usepackage[hidelinks]{hyperref}
\usepackage{svg}
\usepackage{caption}
\usepackage{float}
\usepackage{subcaption}
\usepackage{bbm}

\usepackage{ragged2e}
\usepackage{graphicx}

\newcommand\blfootnote[1]{%
  \begingroup
  \renewcommand\thefootnote{}\footnote{#1}%
  \addtocounter{footnote}{-1}%
  \endgroup
}

\begin{document}

\title{The hardware is the software}

\author{Jérémie~Laydevant}
\email{jl3668@cornell.edu}
\affiliation{School of Applied and Engineering Physics, Cornell University, Ithaca, NY 14853, USA}
\affiliation{USRA Research Institute for Advanced Computer Science, Mountain View, CA 94035, USA}

\author{Logan~G.~Wright}
\email{logan.wright@yale.edu}
\affiliation{School of Applied and Engineering Physics, Cornell University, Ithaca, NY 14853, USA}
\affiliation{NTT Physics and Informatics Laboratories, NTT Research, Inc., Sunnyvale, CA 94085, USA}
\affiliation{Department of Applied Physics, Yale University, New Haven, CT 06511, USA}

\author{Tianyu~Wang}
\email{wangty@bu.edu}
\affiliation{School of Applied and Engineering Physics, Cornell University, Ithaca, NY 14853, USA}

\author{Peter~L.~McMahon}
\email{pmcmahon@cornell.edu}
\affiliation{School of Applied and Engineering Physics, Cornell University, Ithaca, NY 14853, USA}
\affiliation{Kavli Institute at Cornell for Nanoscale Science, Cornell University, Ithaca, NY 14853, USA}

\blfootnote{\hspace{-0.25cm}*,$\dagger$ : These authors contributed equally to this work.\vspace{0.05cm}}

\begin{abstract}
\vspace{0.2cm}
\begin{center}
\textit{This is an opinion paper prepared for the journal Neuron's NeuroViews section.}
\end{center}
\vspace{0.2cm}

Human brains and bodies are not hardware running software: the hardware is the software. We reason that because the microscopic physics of artificial-intelligence hardware and of human biological ``hardware'' is distinct, neuromorphic engineers need to be cautious (and yet also creative) in how we take inspiration from biological intelligence. We should focus primarily on principles and design ideas that respect---and embrace---the underlying hardware physics of non-biological intelligent systems, rather than abstracting it away. We see a major role for neuroscience in neuromorphic computing as identifying the physics-agnostic principles of biological intelligence---that is the principles of biological intelligence that can be gainfully adapted and applied to any physical hardware.

\end{abstract}
\maketitle
\vspace{-0.5cm}

\section{Prologue: Life, with lasers}
On a planet far, far away, intelligent lifeforms known as Lasans think, move, and feel. They may even conduct experiments and write scientific articles. They are, in other words, like humans. Except, however, for one notable detail: the early ancestors of Lasans, flailing their way through the deep, dark oceans of the Lasan planet, evolved the ability to produce and detect laser radiation---which allows them to communicate and process information in the optical domain.

The eery bioluminescence of Earth’s deep-sea creatures may convince you that the Lasan’s evolutionary trajectory is plausible. Our motivation to include Lasans in the opening of this NeuroView is not plausibility however. Rather, it is to pose the question: How would the infrastructure of intelligence (i.e., Lasan ``brains'' and beyond) differ from our own? And in what ways would they be similar?

\begin{center}    
\begin{figure}[h]
\includegraphics[width=\textwidth, trim={0 6cm 0 0},clip]{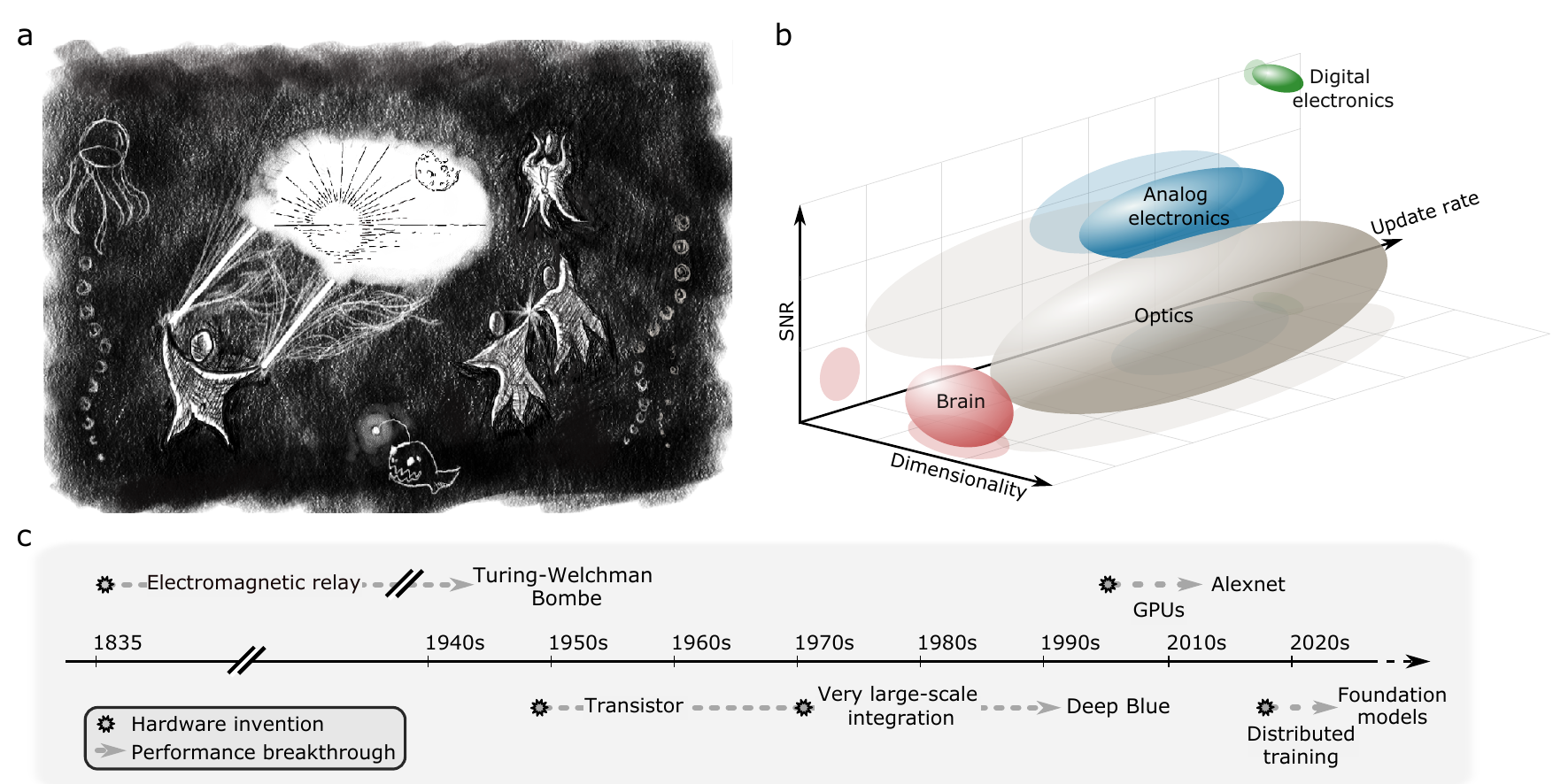}
\caption{\justifying \textbf{The available hardware determines what and how intelligence can be engineered.} \textbf{a}. Deep-sea creatures on a far-away planet think and communicate using light. \textbf{b}. Different physics leads to different attributes of computing hardware, such as the signal-to-noise ratio (SNR) of information-carrying signals; update rate---the speed of the hardware's dynamics, which affects how quickly calculations can be performed; and dimensionality---how many spatial dimensions the hardware can take form in. Different algorithms work formidably well in the brain and on digital electronics because they leverage the hardware at its best. We expect that hardware-agnostic principles of intelligence would also map to unconventional hardware such as optics or analog electronics. \textbf{c}. The hardware-shaped evolution of AI algorithms. Performance breakthroughs on specific tasks have tended to rely on software that efficiently exploits the strengths of the available computing machinery. This observation---that the best algorithms are those that best exploit the available hardware---has been made several times before \cite{rodney1999cambrian,sutton2019bitter,hooker2020lottery}, and described as ``the bitter lesson'' \cite{sutton2019bitter} and ``the hardware lottery'' \cite{hooker2020lottery}.}
\label{fig:fig1}
\end{figure}
\end{center}

It seems easier to answer the first question. As our vast networks of optical telecommunications stand testament to, laser light is an excellent way to efficiently transmit information over long distances---far better than electrical or biochemical transmission. The biological hardware of Lasan intelligence would thus not be as constrained by the costs of communicating information (our brains are very different---there, communication costs account for nearly 35 times more energy consumption than neural computations \cite{levy2021communication}). This capacity for efficient long-range communication means that Lasans would likely evolve to communicate with each other directly by their medium of thought, laser light (or even holographically, as in Figure~\ref{fig:fig1}a). What would this mean for an individual Lasan’s sense of self? Or for inventions like language and writing?

We will halt our speculations here, but suffice to say Lasan ``brains'', not to mention bodies and societies, would be quite different from our own. In what ways would they be similar? We can think of at least two important ways. 

First, the evolution of intelligence in the Lasans’ ancestors would almost certainly follow from the same basic need as it did on Earth: intelligence is required to navigate, to orchestrate motion within a complex, dynamic environment\cite{llinas2002vortex}. In other words, intelligence would evolve to serve the first and perhaps only purpose of \textit{predictive control}: predicting the future in order to move within it. Our earthy biological intelligence first evolved from the need to navigate and control motion while embodied in a dynamic, complex world\footnote{Of course, beyond this initial impetus, our own and other biological intelligences have evolved additional abilities. However, these have come after and from the initial need of predictive control, and biological intelligence has been fundamentally shaped by this order.}.

Second, the structures of Lasan thought would likely emerge from the physics of those thoughts’ implementation. Our brains are the way they are because of the physical constraints of biochemical diffusion, of electricity, of thermodynamics, etc. The intelligence of the Lasans would likewise be shaped by the physics of, among many other things, laser light. As first-year biology students are taught: Form is function. Our brains and bodies are not hardware running software---the hardware is the software.

\section{The Hardware is the Software: Introduction}

The lives and thoughts of Lasans may be imaginary, but the question of how their intelligence would be different and similar to ours is real. Understanding biological intelligence involves understanding not only how and why we humans think and act the way we do, but the same for dogs, for octopi, for ants, and perhaps even Lasans.

What are the principles of biological intelligence that transcend physical implementation? The purpose of this article is \textit{not} to argue that this question is important for neuroscience (we think that most neuroscientists already agree it is!). Rather, the purpose of this article is to argue that this question should (re)define the relationship between neuroscience, neuromorphic engineering, and artificial intelligence (AI). As subsequent sections will develop, our argument is essentially as follows:

\begin{enumerate}
\item The physics of the substrates we construct AI systems from are fundamentally different from the physics of our brains. Although often invisible, this physics (in the form of hardware strengths and constraints) has arguably been the predominant driving force for modern AI. Neuroscience inspirations have been less important (so far).  
\item Modern AI systems have nonetheless been shaped by ideas that were (or at least could have been) learned by studying biological intelligence. Ideas that have had impact have, however, been mostly physically transcendent, not based on, e.g., the details of human cellular biology.
\item If future breakthroughs in AI are to arise from neuroscience (and we think this would be good---it would save a tremendous amount of time, money, and anguish), then neuroscientists must collaborate with AI researchers, computer scientists, physicists, and others in understanding and defining physically transcendent principles of biological intelligence. 
\end{enumerate}

\section{A core theme of neuromorphic computing has been to force non-biological systems to act like biological systems}

As is common among computer scientists of all stripes, many (perhaps even most) AI scientists have avoided thinking too much about hardware, favoring instead a focus on abstracted algorithms. This has, of course, not always been the case \cite{rosenblatt1958,hopfield1984}, and as AI models are being scaled to the limits (and beyond) of current hardware, there has been a growing recognition that it is beneficial to consider the hardware in AI innovation \cite{hooker2020lottery}.

Historically there have been at least three main goals for neuromorphic computing. Some neuromorphic-computing efforts relate to the goal of constructing, at scale, a simulator of the brain, and are motivated by the brain science such a simulator would enable. Other efforts have aimed to mimic the brain not necessarily for ultra-efficient computation, but to interact with sensory data, as in the early efforts to realize analog electronic silicon retinas, cochleas and so on \cite{mead1989vlsi,mead2020how}. Finally, many current efforts are motivated by the desire to create energy-efficient computers, especially for AI \cite{markovic2020physics}.

The silicon retina exemplifies neuromorphic computing in its glory but also in its drifts. The silicon retina was jointly inspired by biology and semiconductor physics---it is relatively easy and even natural to copy the logarithmic response of human retinas in silicon by using transistors. This harmonious translation led to a revolutionary effect: not only did the silicon retina offer enhanced dynamic range, but it also exhibited an emergent biological byproduct, enhancing the edges of the images. It is easy to view this retrospectively as the eureka moment for neuromorphic computing. While the work of Mead and colleagues on the silicon retina is now largely remembered for its biological insight, insight from semiconductor physics was also crucial. Simply copying biology does not necessarily confer an advantage if the facsimile is made on a substrate with fundamentally different physical constraints. A good translator must do more than just convert words, one by one, from one language to another; more nuanced consideration is required, e.g.. of grammar, rhythm, cultural context, etc. A neuromorphic engineer is a translator of physical language---this requires intimate consideration of both the source material (biology) and the adaptation (the new hardware's microscopic physics). It is, after all, the hardware that offers features to exploit, not the inverse.

One important mode in which this sort of ``mistranslation'' has occurred is by translating biology into idealized mathematical models, and then realizing those idealized models in new hardware, whether or not the algorithms implicit to the models are well-suited to the new hardware. Neuromorphic computing has, stubbornly, nearly always forced the hardware implementation to realize abstractions such as the original McCulloch \& Pitts neuron with ideal linear synapses, ideal non-linear neurons (whether spiking or not), ideal Multiply-and-Accumulate (MAC) operations, etc. Implementing these idealized mathematical models or operations frequently requires complicated electronic circuits, operating electronic devices in carefully controlled, constrained regimes, and other engineering trade-offs that reduce efficiency. The translation of biological information processing into mathematics is an important endeavor---one that has been absolutely foundational in both theoretical neuroscience and modern artificial intelligence. But this translation is not the same translation that must be performed for neuromorphic hardware because hardware is not mathematics. In our view, the forced adherence to specific mathematical models has limited the impact of neuromorphic computing.

To be clear, our point is not that emulating biology specifically is wrong, but rather that forcing physical systems to do things that do not come naturally will inevitably reduce performance. This is true when the unnatural behavior is copying biology with electronics, but it is also true for any other idealized function a physical substrate does not natively, simply provide. In our view, the ideal path for neuromorphic computing must involve a similar intellectual exercise as estimating the biological intelligence of extraterrestrials---rather than asking only what biology has done here on Earth, as neuromorphic engineers we must ask what biology would do, given the ``alien'' physics of our favorite hardware. 

\section{Physical constraints define both our brains and modern AI systems}

The brain---a flabby piece of hardware made of water, molecules, ions, etc.---intuitively has fundamentally very different physical constraints than digital processors made of solid-state semiconductors. But how do the microscopic differences lead to macroscopic differences, i.e., differences that would affect the algorithms or software run on the hardware?

Consider this easy example: transmitting information within the hardware from one region to another---a necessity for any information-processing system. While the brain and semiconductor-based processors both encode information mostly with voltages, the underlying particles are different---electrons for semiconductors, and ions and neurotransmitters for the brain. The differences in physics leads to radically different temporal and spatial scales (Figure~\ref{fig:fig1}b).

Hardware has a long history of shaping the software of AI (Figure~\ref{fig:fig1}c). For example, the advantages of digital electronic circuits have defined the battle between machines and humans in chess. Although the eventual success of machines still relied on some human expertise, Deep Blue's algorithm is not what we would today consider a particularly brain-like algorithm. Rather, Deep Blue---based on special-purpose CMOS hardware---competed using an algorithm suited to the strengths of its hardware. Machine success in chess (as well as AI and many, many other domains of modern human activity) has overwhelmingly been achieved through a process of software better-utilizing the compute capabilities of expanding digital electronics, with algorithms that succeed harnessing this compute most effectively \cite{rodney1999cambrian,hooker2020lottery,sutton2019bitter}.

The most recent repeat of this story has been the hardware influence of graphics processing units (GPUs) on AI. It is by now widely appreciated that GPUs are exceptionally well-suited to implementing deep artificial neural networks based on large matrix multiplications, such as multilayer perceptrons, convolutional neural networks, and most recently, Transformers. This insight, exploited by the authors of AlexNet in 2012\cite{LeCun2015}, has had echoes in the recent use of clusters of GPUs (and other specialized neural-network ``accelerators''), which has enabled the scaling of training necessary for today's large language models.

We think it is incredibly unlikely that this same story---called ``the bitter lesson'' by Sutton \cite{sutton2019bitter} and ``the hardware lottery'' by Hooker \cite{hooker2020lottery}---has been told for the last time. Rather, we see this as a defining feature of the evolution of computing.

When examining the state of modern AI it is tempting to conclude that the greatest influence on the form of modern AI algorithms has been the underlying physical constraints of digital integrated circuits, rather than any neuroscience insight. While we think this is at least partially true, its exceptions are notable, and serve as important clues for how neuroscience could help guide a new co-evolution of hardware and software for AI. 

\section{Neural principles have nonetheless shaped modern AI, but only where they are compatible with the underlying constraints of physical hardware.}

Perhaps the clearest influence of neuroscience on modern AI is connectionism. The hypothesis that vast networks of connected neurons can exhibit emergent intelligence is today being increasingly tested. We think it is already fair to conclude that current large-scale artificial neural networks (ANNs) exhibit emergent capabilities (how far this can be pushed remains to be seen). 

Other examples of neuroscience concepts can been seen throughout modern large-scale ANNs\footnote{We don't know if, in each case, the first users of these concepts in AI were explicitly influenced by the neuroscience literature---but in principle they could have been.}, from autoregressive pretraining of large language models \cite{gpt3}, whose next-word prediction resembles the next-sensation predictive objective of our brains; to the concept of pretraining and fine-tuning \cite{gpt3}, which resemble the interplay of evolution, instinct, and learning; to reinforcement learning, which resembles the learning (and teaching) humans experience in early childhood and beyond. We don't know if, in each case, the first users of these concepts in AI were explicitly influenced by the neuroscience literature---but in principle they could have been. These examples illustrate the foundational influence\footnote{Again, we are not making a historical statement about how the inventors of various AI methods and architectures came to the ideas they used. Rather, we are noting that ideas from neuroscience have ultimately also been important in designing AI, possibly through reinvention in some cases.} ideas from neuroscience and psychology have in artificial intelligence. However, these examples also illustrate how the influence has primarily been at a high level: at the level of software, rather than directly guiding the design of AI hardware.

One reason for this is simply that hardware is expensive to experiment with; while relatively small-scale neuromorphic-hardware experiments are plentiful in the research literature, the large-scale, highest-performing hardware that mainstream AI runs on costs hundreds of millions of dollars (or more) to develop, so radical explorations in design are very rare. 

The second reason is our main thesis: the physics of our brains and bodies are very different from those of any substrate we are likely to use for present or near-future computers, including neuromorphic computers, so for neuroscience insight to impact how state-of-the-art AI is done, it cannot be too specific to the physical details of the implementation in animals. The past decade has seen billions of dollars of investment in the development of AI-specific chips, including some with novel physical substrates beyond standard CMOS electronics, and it seems likely that there will continue to be large investment in new hardware for AI. However, this new hardware will not change our thesis: the new hardware will most likely still be based on integrated electronics. Photonics may play a role, but as our introductory thought experiment suggested, this would push AI hardware physics even further away from that of biological brains.

In summary, insights from neuroscience will be most impactful in AI if they are physics-agnostic or are particularly amenable to implementation with semiconductor electronics and photonics, rather than biological matter. AI has already greatly benefited from discoveries in neuroscience about the basis of biological intelligence, and we believe that there is much more for AI researchers to adapt from neuroscience, but it is useful to keep in mind that the physics of the hardware affects which and how different insights can and should be applied.

\section{Conclusions and looking forward}
The rapid expansion of both artificial intelligence (AI) and neuroscience means that, perhaps more than ever, neuroscientists have a valuable role to play in informing the design of synthetic-intelligence systems. This includes both the software as well the hardware. However, the physical capabilities and constraints of the hardware substrates used to implement modern AI systems are not the same as those of the biological substrates that make up biological intelligence. Analogously to how form determines function throughout biology, hardware specifics---physics and engineering design---have constrained the shape of AI software. While the physics of electronics is immutable, the design of electronic hardware is not, and there is growing interest in engineering special-purpose hardware for AI.  

We think that a key role neuroscientists can play in the next phase of AI software and hardware development is related to the question posed in our introduction: what would biological intelligence look like if it were based on very different microscopic physics than that of our human brains? Which aspects would remain the same, and which aspects would be very different? Answering these questions involves both identifying universal, physics-agnostic principles of biological intelligence, and a deeper understanding of how our own biological intelligence, and silicon-electronics-based artificial intelligence, are shaped by their respective hardware physics.

Our concluding questions, which we---as non-neuroscientists\footnote{All the authors of this article are engineering physicists who are interested in AI hardware and software.}---think neuroscientists could help answer to our great benefit, are: 
\begin{itemize}
    \item What are the universal, physics-agnostic principles of biological intelligence?
    \item Which of these principles do you think have been stubbornly ignored by earlier generations of artificial intelligence? Or, if not \textit{ignored}, then underexplored and worth revisiting? (Either at the software level, the hardware level, or---ideally---both.)
    \item Conversely, which principles of Earthly biological intelligence are likely to be specific to our specific biology (and therefore of questionable use in hardware based on very different microscopic physics)?
\end{itemize}

We hope that this essay stimulates further discussion and engagement between the neuroscience and neuromorphic-computing communities. We encourage you to get in touch with us (even, or perhaps especially, if you do not agree with all we have said here!). Our plan is to eventually follow this short NeuroView article with another piece, one that will attempt to summarize your and our answers. We would be delighted to have you join us in conversation, as a co-author, and/or in friendly debate, helping to define both new directions and new questions for the intertwined futures of AI, neuromorphic computing, and neuroscience.

\section*{Acknowledgements}
We wish to thank NTT Research for their financial and technical support (P.L.M., T.W. and L.G.W.). This work has also been supported by the National Science Foundation (award no. CCF-1918549: P.L.M., J.L., and T.W.; award no. CBET-2123862: P.L.M.) and a David and Lucile Packard Foundation Fellowship (P.L.M.). T.W. acknowledges partial support from an Eric and Wendy Schmidt AI in Science Postdoctoral Fellowship. We thank Gautam Reddy for helpful feedback on a draft of our manuscript.

\section*{Author Contributions}
J.L. and L.G.W. outlined the manuscript and wrote the initial draft with feedback from P.L.M. and T.W.. J.L. and T.W. created the figures. P.L.M. supervised the team and edited the final draft with input from all authors.

\section*{Declaration of Interests}
Subsets of the authors are listed as inventors on patent filings related to neuromorphic computing, as follows. L.G.W., T.W. and P.L.M. are listed as inventors on U.S. provisional patent application \#63/392,042. L.G.W. and P.L.M. are listed as inventors on U.S. provisional patent application \#3/178,318. T.W. and P.L.M. are listed as inventors on U.S. provisional patent application \#63/149,974.

\bibliography{references}
\end{document}